\documentclass{article} 
\usepackage[final]{colm2026_conference}

\usepackage{microtype}
\usepackage{graphicx}
\usepackage{subcaption}
\usepackage{booktabs} 

\usepackage{hyperref}       
\usepackage{url}            
\usepackage{booktabs}       
\usepackage{amsfonts}       
\usepackage{nicefrac}       
\usepackage{microtype}      
\usepackage{xcolor}         
\usepackage{tabularx,booktabs,array}
\newcolumntype{Y}{>{\centering\arraybackslash}X}
\usepackage{xspace}
\usepackage{enumitem}
\usepackage{amsmath}

\usepackage{booktabs}
\usepackage{multirow}
\usepackage{colortbl}       
\usepackage[table]{xcolor}  
\usepackage{graphicx}
\usepackage{wrapfig}
\usepackage{stmaryrd}
\usepackage{listings}
\usepackage{tcolorbox}
\usepackage{amssymb,amsthm}
\usepackage{mathpartir}
\usepackage{mathtools}
\usepackage{cleveref}
\usepackage{pifont}

\newcommand{\cmark}{\ding{51}}
\newcommand{\xmark}{\ding{55}}\usepackage{pifont}

\newcommand{\name}{Quokka\xspace}

\newcommand{\evalsize}{866\xspace}    

\newcommand{\trainfinalsize}{3589\xspace} 

\newcommand{\uauto}{UAutomizer\xspace}

\newcommand{\CodeIn}[1]{\texttt{#1}\xspace}

\newcommand{\numllm}{9\xspace}

\usepackage{lineno}

\definecolor{darkblue}{rgb}{0, 0, 0.5}
\hypersetup{colorlinks=true, citecolor=darkblue, linkcolor=darkblue, urlcolor=darkblue}

\usepackage{amssymb}
\usepackage{mathtools}
\usepackage{amsthm}

\title{Quokka: Accelerating Program Verification with LLMs via Invariant Synthesis}

\author{
\textbf{Anjiang Wei}\textsuperscript{1},
\textbf{Tianran Sun}\textsuperscript{2},
\textbf{Tarun Suresh}\textsuperscript{3},
\textbf{Haoze Wu}\textsuperscript{4},
\textbf{Ke Wang}\textsuperscript{5},
\textbf{Alex Aiken}\textsuperscript{1}
\\
\textsuperscript{1}Stanford University \hspace{1em}
\textsuperscript{2}Shanghai Jiao Tong University \hspace{1em}\\
\textsuperscript{3}University of Illinois Urbana-Champaign \hspace{1em}
\textsuperscript{4}Amherst College \hspace{1em}
\textsuperscript{5}Nanjing University
}

\begin{document}

\ifcolmsubmission
\linenumbers
\fi

\maketitle

\begin{abstract}
Program verification relies on loop invariants, yet automatically discovering strong invariants remains a long-standing challenge. We investigate whether large language models (LLMs) can accelerate program verification by generating useful loop invariants. We introduce \name{}, a framework for LLM-based invariant synthesis with soundness guarantees and state-of-the-art performance. Unlike prior work that treats LLM outputs as noisy symbolic material requiring substantial post-processing, \name{} adopts a simpler algorithm design that directly validates whether each LLM-generated invariant helps prove the target assertion. We construct a benchmark of \evalsize{} evaluation instances and \trainfinalsize{} training instances derived from SV-COMP, and evaluate \numllm{} LLMs spanning multiple model families. We demonstrate that supervised fine-tuning and Best-of-N sampling yield measurable improvements, and we show that \name{} consistently outperforms prior LLM-based verifiers. Our code and data are publicly available at \url{https://github.com/Anjiang-Wei/Quokka}
\end{abstract}

\section{Introduction}
\label{sec:intro}

Program verification aims to provide formal guarantees that software behaves as intended, with applications in many safety-critical domains~\citep{fan2017dryvr,luckcuck2019formal}. A long-standing challenge in this area, studied for more than four decades, is the automatic discovery of loop invariants. In this work, we investigate whether large language models (LLMs) can accelerate program verification by generating useful loop invariants.

\begin{figure*}[!tb]
    \centering
    \includegraphics[width=\linewidth]{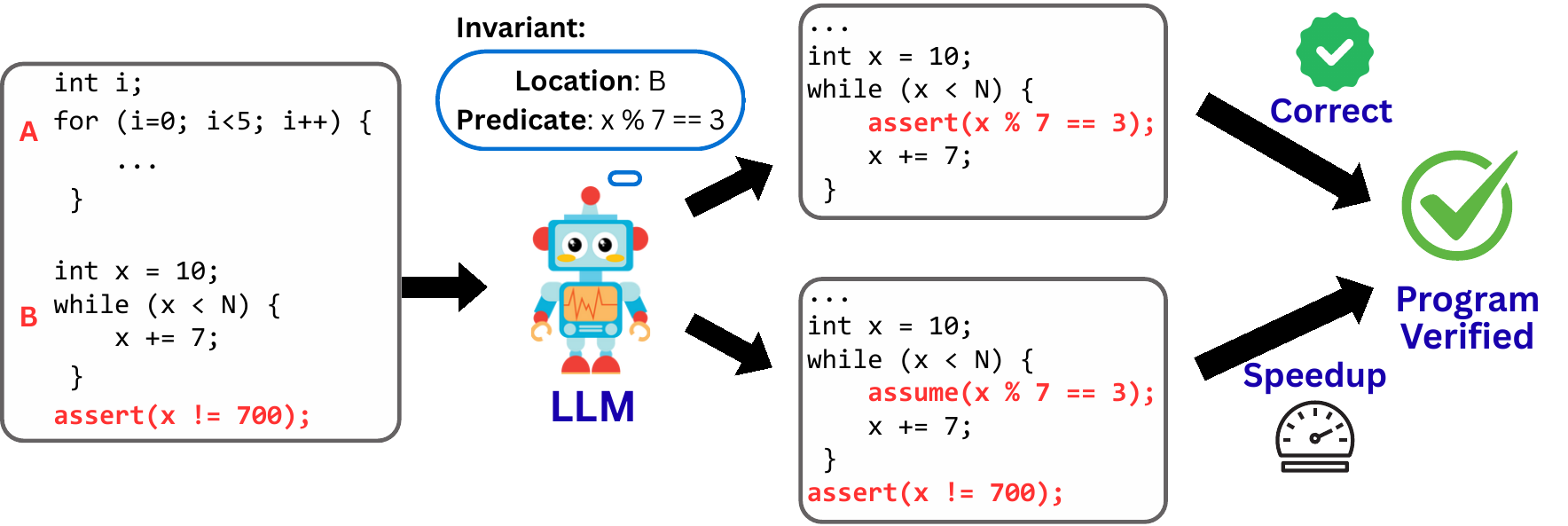}
    \caption{Illustration of \name{}'s evaluation pipeline. The LLM proposes an invariant by specifying a program location and predicate (e.g., location B with \texttt{x \% 7 == 3}). The verification algorithm then incorporates this invariant to prove the property \texttt{x != 700} using two verifier queries, and we measure the resulting speedup relative to a baseline without LLM assistance.}
    \label{fig:intro}
\end{figure*}

Loop invariants are conditions that hold before and after each loop iteration, and they are central to deductive program verification. To accelerate program verification, loop invariants must not only be correct but also sufficiently strong to prove the assertions. Generating correct invariants is relatively easy, since any universally true condition qualifies. However, only strong invariants can reduce verification effort and lead to a speedup. For example, in \Cref{fig:intro}, the invariant $x > 0$ is correct but not strong enough to prove the final assertion $x \neq 700$, whereas $x \equiv 3 \pmod{7}$ is both correct and sufficiently strong.

Discovering such invariants is difficult and undecidable in general, which has motivated a long line of research. Traditional approaches include constraint solving~\citep{solving2003linear,gupta2009tests}, dynamic analysis~\citep{le2019sling}, etc. Since invariant discovery is challenging, researchers have tried a variety of learning-based methods~\citep{li2017automatic,ezudheen2018horn}. Building on this progression, the strong capabilities of LLMs in code generation and program reasoning~\citep{austin2021program,chen2021evaluating,wei2025codearc} naturally motivate a systematic evaluation of their potential for invariant discovery.

The first work~\citep{pmlr-v202-pei23a} to evaluate LLMs for invariant generation has two key limitations. One limitation is that it only checks correctness without assessing whether invariants are strong enough to accelerate verification. Another limitation is that correctness is determined by directly comparing against Daikon~\citep{ernst2007daikon}, a dynamic analysis tool that infers invariants from test executions rather than formal verification. This approach is unsound: Daikon's invariants may themselves be incorrect. Moreover, semantically equivalent invariants (e.g., $a > 0$ vs. $a \ge 1$ for integers) can be incorrectly rejected as wrong. Consequently, prior work cannot reliably evaluate either the correctness or the practical utility of LLM-generated invariants.

A series of follow-up works have proposed LLM-based verifiers. Instead of evaluating LLMs in isolation, these efforts develop verification frameworks powered by LLMs. This line of work predominantly adopts complex algorithms to handle LLM-generated invariants, treating LLM output as noisy symbolic material that must be algorithmically repaired or reassembled. For instance, LaM4Inv~\citep{wu2024llm} uses a ``query-filter-reassemble'' strategy that filters and combines model-generated predicates through complex logical operations (conjunctions or disjunctions) to construct valid loop invariants, rather than using them directly. Clause2Inv~\citep{cao2025clause2inv} extends this idea further. LOOPY~\citep{kamath2024leveraging} employs a customized Houdini algorithm with iterative refinement to filter candidate invariants. LEMUR~\citep{wu2024lemur} develops a recursive and backtracking algorithm for fixing invariants. While these approaches demonstrate the potential of LLM-based verification, their complexity raises a fundamental question: as models become increasingly capable, do we really need such elaborate, harness-heavy designs for building LLM-based verifiers? What if we stop rescuing LLM-generated invariants and instead measure directly whether a candidate invariant helps the solver?

In this work, we introduce \name{}, a framework for LLM-based invariant synthesis with soundness guarantees and state-of-the-art performance. Prior LLM-based verification systems treat LLM outputs as imperfect intermediate artifacts requiring substantial symbolic post-processing—predicate filtering, clause combination, Houdini pruning, repair, or backtracking. In contrast, \name{} adopts a simpler design: it directly evaluates LLM-generated invariants by checking whether each invariant is valid and whether it helps prove the target assertion. This design eliminates most customized post-processing logic, enables parallel verification queries, and shifts the objective from merely finding correct invariants to finding invariants that can accelerate verification. Despite its simplicity, \name{} achieves state-of-the-art performance compared to all prior LLM-based verifiers.

To support systematic comparison across solvers and LLMs, we construct a dataset of \evalsize{} instances derived from the most recent edition of the software verification competition SV-COMP~\citep{beyer2025improvements}. To the best of our knowledge, this is the largest evaluation dataset for LLM-based verifiers to date. After evaluating \numllm{} LLMs, we find that \name{} proposes a meaningful and challenging setup for LLMs, where model capabilities can be effectively distinguished, and there remains significant room for improvement on challenging cases.

Beyond model evaluation, we investigate techniques to enhance invariant generation capabilities. We construct a training dataset of \trainfinalsize{} instances using a verifier-based filtering approach that ensures data quality. Our experiments demonstrate that both supervised fine-tuning and Best-of-N sampling yield measurable improvements in accelerating verification.

In summary, our contributions are as follows:
\begin{itemize}
    \item We introduce a method for LLM-based invariant synthesis that provides sound evaluation and achieves the best performance results compared with prior LLM-based verifiers.
    \item We construct a benchmark of \evalsize{} instances, and conduct an evaluation of \numllm{} LLMs across multiple model families.
    \item We develop a new verifier-guided data generation method for constructing training data and demonstrate that both supervised fine-tuning and Best-of-N sampling yield measurable improvements in accelerating verification.
\end{itemize}

\section{Related Work}
\label{sec:related}

\paragraph{Traditional Methods for Program Invariant Generation.}
A long line of research has explored invariant synthesis using traditional techniques without machine learning, including model checking~\citep{flanagan2002predicate,lahiri2007predicate,hojjat2018eldarica,vediramana2024global}, abstract interpretation~\citep{karr1976affine,cousot1977abstract,cousot1978automatic,cousot1979systematic}, constraint solving~\citep{gulwani2009constraint,gupta2009tests}, Craig interpolation~\citep{jhala2006practical,mcmillan2010lazy}, and syntax-guided synthesis~\citep{fedyukovich2018accelerating}. Prior work evaluating LLM-generated invariants~\citep{pmlr-v202-pei23a} has relied on Daikon~\citep{ernst2007daikon}, a tool for dynamic invariant detection~\citep{echenim2019ilinva,le2019sling}. Daikon executes the program, observes runtime values, and reports properties that consistently hold over the observed executions. However, such invariants may fail to generalize to all possible executions, thereby compromising soundness. Our approach instead employs a verifier-based algorithm relying on \uauto~\citep{schussele2024ultimate} that ensures soundness.

\paragraph{Learning-Based Method for Invariant Generation.}
Machine learning based techniques have been widely adopted in invariant synthesis, including decision tree~\citep{garg2014ice,garg2016learning,ezudheen2018horn,riley2022multi,xu2020interval}, support vector machine~\citep{li2017automatic,sharma2012interpolants}, reinforcement learning~\citep{si2018learning,yu2023loop}, and others~\citep{sharma2013verification,ryan2019cln2inv,10.1145/3385412.3385986}. More recently, large language models have demonstrated strong capabilities in reasoning about code and logic~\citep{wei2025equibench,wei2025codearc,wei2025satbench}, motivating a growing body of work on using LLMs for invariant synthesis. \cite{pmlr-v202-pei23a} is the first work to systematically evaluate LLMs' capabilities in generating invariants, although its evaluation does not provide formal soundness guarantees. Subsequent work has explored various ways of coupling LLMs with symbolic solvers, including ranking LLM-generated invariants~\citep{chakraborty2023ranking}, the ``query-filter-reassemble'' strategy of LaM4Inv~\citep{wu2024llm}, the recursive backtracking procedure of LEMUR~\citep{wu2024lemur}, and Loopy's integration of the classic Houdini algorithm~\citep{kamath2024leveraging}. On the dataset side, \cite{liu2024towards} introduces a rule-based method for constructing a fine-tuning corpus, in contrast to our verifier-based data generation approach. Among these methods, LEMUR and Loopy are particularly relevant to our setting but differ substantially in their verification procedures. LEMUR uses a recursive, sequential, and backtracking-based algorithm in which proving an invariant can itself introduce a new proof goal and recursively invoke the LLM. Although LEMUR and Quokka share the same underlying solver, Quokka intentionally adopts a substantially simpler algorithm. Loopy instead builds on Frama-C and combines an adaptation of the Houdini algorithm for filtering incorrect LLM outputs with LLM-based repair of candidate invariants. In contrast, our work develops a simple and sound evaluation algorithm for assessing LLM-generated invariants and investigates how fine-tuning and Best-of-N sampling can improve LLM performance in invariant synthesis.
\section{Method}
\label{sec:method}

\subsection{Preliminary}
\label{subsec:prelim}

We formalize the task of loop invariant synthesis using standard Hoare logic~\citep{hoare1969axiomatic}. A Hoare triple $\{P\}\, S \,\{Q\}$ specifies that if the precondition $P$ holds before executing a statement $S$, then the postcondition $Q$ will hold after its execution. In the context of loops, an invariant $I$ is a logical proposition that summarizes the state of the program at each iteration, and it is the key to proving the validity of Hoare triples involving loops. For a loop of the form \texttt{while} $B$ \texttt{do} $S$, the goal of invariant synthesis is to identify a loop invariant $I$ that satisfies the following inference rule:

\[
\frac{P \Rightarrow I \quad \{I \wedge B\}\, S \,\{I\} \quad I \wedge \lnot B \Rightarrow Q}
{\{P\}\ \texttt{while}\ B\ \texttt{do}\ S\ \{Q\}}
\]

Here, $P$ is the precondition, $Q$ is the postcondition, $B$ is the loop condition, and $S$ is the loop body. 
Intuitively, the inference rule requires that the invariant $I$ holds at the beginning of the loop ($P \Rightarrow I$), is preserved across every iteration of the loop body ($\{I \wedge B\}\, S \,\{I\}$), and upon termination ensures the postcondition ($I \wedge \lnot B \Rightarrow Q$). 

Invariant synthesis amounts to generating a logical summary $I$ that is both \emph{correct}, meaning it can be verified, and \emph{strong}, meaning it enables verification of the final assertion. Weak but correct invariants contribute little, leaving most of the reasoning to the verifier, whereas strong invariants narrow the search space of program states, reduce solver effort, and yield substantial speedups.

\newcommand{\True}{\ensuremath{\mathbf{T}}}
\newcommand{\False}{\ensuremath{\mathbf{F}}}
\newcommand{\Unk}{\ensuremath{\mathbf{U}}}
\newtheorem*{theorem*}{Theorem}
\newtheorem*{lemma*}{Lemma}

\newcommand{\Prog}{\mathcal{P}}

\subsection{Verifier-Based Approach}
\label{subsec:decision}

We formalize our verifier-based algorithm for assessing candidate invariants.  
Let \(\Prog\) denote a program, which may contain multiple loops and functions. A \emph{property} is written as \(p=\langle \varphi,\ell\rangle\), where \(\varphi\) is a state predicate and \(\ell\) is a program location. For a finite set \(A\) of properties, let \(\mathrm{Asm}(\Prog,A)\) be the program obtained from \(\Prog\) by inserting \texttt{assume} statements for all elements of \(A\). An execution of \(\mathrm{Asm}(\Prog,A)\) that reaches a location where an assumption is violated terminates immediately. We write \(\Prog \models_{A} p\) to indicate that all executions of \(\mathrm{Asm}(\Prog,A)\) satisfy the assertion \(p\). The notation \(\Prog \models p\) abbreviates \(\Prog \models_{\varnothing} p\). Since assumptions restrict behaviors, if \(\Prog \not\models_{A} p\) for some \(A\), then necessarily \(\Prog \not\models p\).

We assume access to a verifier
\[
V(\Prog,A,p) \in \{\True,\False,\Unk\},
\]
which returns either \True\ (proved), \False\ (refuted), or \Unk\ (inconclusive).  
The verifier is required to be sound on conclusive outcomes:
\[
V(\Prog,A,p)=\True \;\Rightarrow\; \Prog \models_{A} p,
\]
\[
V(\Prog,A,p)=\False \;\Rightarrow\; \Prog \not\models_{A} p.
\]
No completeness is assumed for \Unk, which may arise from timeouts or incompleteness of the underlying verifier.

The verification task defines a target property \(p^\star=\langle \varphi^\star,\ell^\star\rangle\). In the example in \Cref{fig:intro}, \(p^\star\) corresponds to the final assertion \CodeIn{x != 700} at its program location.

Given \(\Prog\) and \(p^\star\), a large language model is instructed to propose a \emph{candidate invariant} \(q=\langle \psi,\ell\rangle\), typically at a loop header. In our implementation, we consider loop headers as the only eligible locations for candidate invariants. For programs containing multiple loops, including nested loops or loops in different functions, the LLM is presented with all loop headers as candidate locations, and it generates a single candidate invariant for the entire program. To evaluate the utility of \(q\), the algorithm issues two verifier queries:

\[
d_a := V(\Prog,\varnothing,q)
\]
\[
d_b := V(\Prog,\{q\},p^\star)
\]

The query $d_a$ verifies whether \(q\) is a correct invariant, while the query $d_b$ determines whether the target property holds under the assumption that \(q\) is true. An illustration of these two verifier queries is provided in \Cref{fig:intro}: $d_a$ corresponds to \CodeIn{assert(x \% 7 == 3)} (upper subfigure), and $d_b$ corresponds to verifying the final assertion with the assume statement inserted (lower subfigure). The outcome of the algorithm is expressed as a judgment
\[
\Prog \;\Rightarrow\; \langle p^\star,q\rangle \;\Downarrow\; d
\qquad\text{with}\quad d \in \{\True,\False,\Unk\}.
\]
The interpretation is as follows: if the judgment yields \True, then \(p^\star\) is established on \(P\); if it yields \False, then \(p^\star\) is refuted; and if it yields \Unk, the attempt is inconclusive.

The inference rules defining this judgment are given below. Each rule specifies one possible derivation of the outcome, depending only on the responses of the verifier.

\begin{mathpar}
\inferrule*[right=(DEC\mbox{-}FALSE)]
  { V(\Prog,\{q\},p^\star)=\False }
  { \Prog \;\Rightarrow\; \langle p^\star,q\rangle \;\Downarrow\; \False }
\\
\inferrule*[right=(DEC\mbox{-}PROP)]
  { V(\Prog,\varnothing,q)=\True \\ V(\Prog,\{q\},p^\star)=d \\ d\neq\False }
  { \Prog \;\Rightarrow\; \langle p^\star,q\rangle \;\Downarrow\; d }
\\
\inferrule*[right=(DEC\mbox{-}U)]
  { V(\Prog,\varnothing,q)\neq\True \\ V(\Prog,\{q\},p^\star)\neq\False }
  { \Prog \;\Rightarrow\; \langle p^\star,q\rangle \;\Downarrow\; \Unk }
\end{mathpar}

Rule \textsc{DEC-FALSE} captures short-circuit refutation: if the goal fails even in the restricted program \(\mathrm{Asm}(\Prog,\{q\})\), then it is false on the original program \(\Prog\).
Rule \textsc{DEC-PROP} implements the prove-then-use strategy: once the candidate invariant \(q\) is established, the outcome is exactly the verifier’s answer on the goal under the assumption \(q\), restricted to \(d\in\{\True,\Unk\}\) so as not to overlap with \textsc{DEC-FALSE}.
Rule \textsc{DEC-U} gives explicit conditions for inconclusiveness: the goal is not refuted under \(q\) and \(q\) is not established as an invariant.

\begin{theorem*}[Decision Soundness]\label{thm:decision-soundness}
If \(\Prog \Rightarrow \langle p^\star,q\rangle \Downarrow \True\) is derivable, then \(P \models p^\star\).
If \(\Prog \Rightarrow \langle p^\star,q\rangle \Downarrow \False\) is derivable, then \(P \not\models p^\star\).
\end{theorem*}

\begin{proof}
  For the \(\True\) case, the final rule must be \textsc{DEC-PROP} with \(d=\True\).
  The premises yield \(V(\Prog,\varnothing,q)=\True\) and \(V(\Prog,\{q\},p^\star)=\True\).
  By verifier soundness, \(\Prog\models q\) and \(\Prog\models_{\{q\}} p^\star\).
  Since \(q\) holds on all executions of \(\Prog\), introducing the assumption \(\{q\}\) does not remove executions relevant to \(p^\star\); thus \(\Prog\models p^\star\). For the \(\False\) case, the final rule must be \textsc{DEC-FALSE}. Its premise \(V(\Prog,\{q\},p^\star)=\False\) implies \(\Prog \not\models_{\{q\}} p^\star\) by soundness. Assumptions restrict behaviors; hence, a violation under assumptions entails a violation without them, yielding \(\Prog\not\models p^\star\).
\end{proof}

This theorem establishes that whenever the calculus derives a conclusive outcome, that outcome is correct. The inconclusive case \Unk\ is deliberately conservative: it makes no claim about the truth or falsity of the property and safely captures verifier incompleteness or timeouts.

\subsection{Implementation}
\label{subsec:impl}

We describe the implementation of our verifier-based evaluation framework.  
Given a program \(P\) and target property \(p^\star\), the system evaluates the generated invariants according to the approach. Each invariant \(q=\langle \psi,\ell\rangle\) consists of a program location \(\ell\) and predicate \(\psi\).

\paragraph{Syntactic Validation.}
Before invoking the verifier, we apply a blacklist-style syntactic filter to the generated predicate \(\psi\). We reject state-modifying expressions containing \CodeIn{++}, \CodeIn{--}, compound assignments such as \CodeIn{+=}, or a standalone assignment operator \CodeIn{=}, while allowing standard Boolean and arithmetic operators. We also require the model output to match the expected insertion format and to reference a valid candidate location. Other malformed or semantically invalid expressions are rejected downstream by the C parser or verifier.

\textbf{Parallel Verifier Queries.}
For each candidate \(q\), the algorithm issues two verifier queries, namely \(d_a = V(\Prog,\varnothing,q)\) to check whether \(q\) is an invariant and \(d_b = V(\Prog,\{q\},p^\star)\) to check whether the target holds under the assumption \(q\). These queries are executed in parallel in our implementation, which reduces latency and enables speedup when the proposed invariant is useful for verification. The final outcome is then derived exactly according to the calculus.

\subsection{Supervised Fine-Tuning and Best-of-N Sampling}
\label{subsec:sftdata}

\paragraph{Synthetic Dataset Generation.}
We prompt GPT-4o (template in Appendix~\ref{subsec:prompt-template}) with three seed programs to synthesize new C programs containing loops and assertions. Seed programs are randomly drawn from SV-COMP~\citep{beyer2025improvements}, disjoint from our evaluation set to prevent data leakage. Although SV-COMP programs may have appeared in LLM pretraining data, the public benchmark does not provide ground-truth invariants, so such exposure does not directly reveal the target invariants evaluated in our study. Generated programs may fail to compile, contain invalid assertions, or include multiple assertions. We split multi-assertion programs into separate instances, each with a single assertion and all loops. We then run \uauto{} on each program and discard non-compilable instances or those with invalid assertions, yielding \trainfinalsize{} synthetic programs.

\paragraph{Extract Invariants Generated from \uauto{}.} We extract loop invariants from \uauto{}'s output when proving assertions in the synthetic programs. Each invariant includes its program location and predicate. Since programs may contain multiple loops, we pair each program with each of its loop invariants to form our training dataset. An example is shown in Appendix~\ref{subsec:sftexample}. We perform supervised fine-tuning using LoRA~\citep{hu2022lora} on these ground-truth invariants.

\paragraph{Best-of-N Sampling.} Best-of-N sampling generates multiple candidates and selects the most effective one~\citep{ehrlich2025codemonkeys}. We choose the invariant yielding the largest speedup. Since each candidate requires two parallel verifier queries (\Cref{subsec:decision}), Best-of-N evaluates 2N queries concurrently.
\section{Experimental Setup}
\label{sec:setup}

\paragraph{Dataset.}
We construct our evaluation benchmark from SV-COMP~\citep{beyer2025improvements}, a standard software verification competition, focusing on problems requiring loop invariant synthesis. By filtering instances containing loop keywords (\CodeIn{while} or \CodeIn{for}), we obtain \evalsize{} instances. To the best of our knowledge, this is the largest dataset used in prior LLM-based verifier evaluations. As shown in \Cref{tab:dataset}, our programs are significantly longer than prior work. Manual inspection reveals our dataset contains multiple loops, functions, arrays, and pointers (features largely absent from prior datasets), making it a more challenging benchmark that better distinguishes solver performance. For training data, we construct \trainfinalsize instances with synthetic data generation filtered by the solvers, as described in \Cref{subsec:sftdata}.

\begin{table}[t]
\centering
\small
\setlength{\tabcolsep}{1.5pt}
\renewcommand{\arraystretch}{1.15}
\begin{tabular}{lcccccc}
\toprule
\textbf{Dataset} & \textbf{Avg. \#Lines} & \textbf{\#Instances} & \textbf{Multi-Loops} & \textbf{Multi-Functions} &  \textbf{Arrays/Pointers} \\
\midrule
LaM4Inv~\citep{wu2024llm} & 18 & 316 & \xmark & \xmark & \xmark \\
Clause2Inv~\citep{cao2025clause2inv} & 18 & 366 & \xmark & \xmark & \xmark \\
LEMUR~\citep{wu2024lemur} & 20 & 180 & \cmark & \xmark & \cmark \\
Loopy~\citep{kamath2024leveraging} & 25 & 638 & \cmark & \cmark & \cmark \\
\textbf{Quokka (Ours)} & 52 & 866 & \cmark & \cmark & \cmark \\
\bottomrule
\end{tabular}
  \caption{Dataset statistics.}
  \label{tab:dataset}
\end{table}

\paragraph{Metrics.}
We evaluate LLMs on two dimensions: (1) correctness of generated invariants and (2) performance improvements. Correctness is determined by whether the solver verifies the invariant within a timeout of \(1.2\times\) the original solving time, providing slack to reduce inconclusive outcomes. We report the number of additionally solved instances (unsolvable by the base solver), total solved instances, and average solving time. All measurements represent end-to-end execution time, including LLM inference overhead.

\paragraph{Choice of Solvers.} We use \uauto~\citep{schussele2024ultimate}, the state-of-the-art non-LLM-based solver, as our base solver. Our methodology requires that the solver can verify programs without externally-provided invariants, enabling speedup measurement relative to a functional baseline. We select \uauto over alternatives like Frama-C~\citep{cuoq2012frama} and ESBMC~\citep{gadelha2018esbmc} because it represents the current state-of-the-art (outperforming ESBMC on SV-COMP benchmarks) and, unlike Frama-C, has built-in invariant synthesis capabilities for autonomous verification. To demonstrate generality, we also present ESBMC experiments in Appendix~\ref{subsec:esbmc}.

Additional details such as models and hardware setup are provided in Appendix~\ref{subsec:app:setup}.
\section{Results}
\label{sec:result}

\subsection{Results of LLMs}
\begin{table*}[!tb]
  \centering
  \small
  \renewcommand{\arraystretch}{1.15}
  \setlength{\tabcolsep}{5pt}
  \begin{tabular}{lccccccc}
    \toprule
    \multirow{2}{*}{\textbf{Model}} 
      & \multirow{2}{*}{\textbf{\# Verified Invariants}} 
      & \multicolumn{3}{c}{\textbf{60s}} 
      & \multicolumn{3}{c}{\textbf{500s}} \\
    \cmidrule(lr){3-5} \cmidrule(lr){6-8}
      & 
      & $\Delta$ 
      & \# Solved 
      & $\overline{T}$ (s) 
      & $\Delta$ 
      & \# Solved 
      & $\overline{T}$  (s) \\
    \midrule
    Llama-3.1-8B & 342 & 5 & 321 & 41.9 & 0 & 342 & 309.9 \\
     Qwen2.5-7B & 419 & 5 & 402 & 37.6 & 0 & 417 & 267.5 \\
    claude-opus-4.1 & 487 & 13 & 466 & 34.2 & 1 & 487 & 229.4 \\
    Qwen2.5-72B & 500 & 11 & 474 & 33.8 & 0 & 500 & 221.7 \\
    o3 & 550 & 17 & 534 & 35.9 & 1 & 549 & 198.3 \\
    claude-sonnet-4 & 620 & 16 & 591 & 27.8 & 1 & 620 & 156.0 \\
    claude-opus-4.5 & 689 & 17 & 659 & 24.2 & 1 & 689 & 118.8 \\
    gpt-5.1 & 694 & 15 & 661 & 23.3 & 1 & 694 & 113.8 \\
    gpt-5.2 & 710 & 21 & 681 & 22.2 & 1 & 710 & 105.1 \\
    \bottomrule
  \end{tabular}
  \caption{Performance comparison of LLMs on invariant generation: number of verified invariants, additional instances solved beyond \uauto~baseline ($\Delta$), total solved instances, and average solving time ($\overline{T}$) under 60s and 500s timeouts.}
  \label{tab:llm}
\end{table*}

\begin{table*}[!tb]
  \centering
  \small
  \renewcommand{\arraystretch}{1.2}
  \setlength{\tabcolsep}{5pt}
  \begin{tabular}{lccccccc}
    \toprule
    \multirow{2}{*}{\textbf{Model}} & \multirow{2}{*}{\textbf{\# Verified Invariants}} & \multicolumn{3}{c}{\textbf{60s}} & \multicolumn{3}{c}{\textbf{500s}} \\
    \cmidrule(lr){3-5} \cmidrule(lr){6-8}
    & & \# $\Delta$  & \# Solved & $\overline{T}$ & \# $\Delta$  & \# Solved & $\overline{T}$ \\
    \midrule
    Qwen2.5-7B (base) & 419 & 5 & 402 & 37.6 & 0 & 417 & 267.5 \\
    Qwen2.5-7B (finetuned) & 431 & 7 & 414 & 36.8 & 0 & 429 & 260.8 \\
    \bottomrule
  \end{tabular}
    \caption{Performance comparison of base and fine-tuned Qwen2.5-7B. Fine-tuning leads to modest improvements in both correctness and speedup metrics.}
   \label{tab:sft}
\end{table*}

\Cref{tab:llm} presents the performance of different LLMs when using \name{}. Overall, the table demonstrates substantial variation in both correctness and performance across different models, with larger and more recent models generally achieving better results.

\paragraph{\name{} proposes a meaningful and challenging problem setup for LLMs.} 
The results in \Cref{tab:llm} reveal an interesting performance pattern across timeout budgets. Under the 60s timeout, LLMs solve much more instances beyond the \uauto~baseline (ranging from 5 additional instances for smaller models to 21 for gpt-5.2), demonstrating their ability to accelerate verification on moderately difficult problems. However, under the 500s timeout, the incremental benefit diminishes significantly, indicating that LLMs struggle to provide meaningful invariants for the hardest problems that require more solving time. This pattern suggests that \name{} establishes a well-calibrated benchmark: LLMs already provide meaningful acceleration on a portion of instances, yet significant room remains for improvement on challenging cases. The substantial variation in verified invariants (342 for Llama-3.1-8B to 710 for gpt-5.2) and average solving times (41.9s to 22.2s under 60s timeout) further demonstrates that the benchmark effectively distinguishes model capabilities. While alternative verification algorithms might yield different absolute performance numbers, the superiority and simplicity of \name{} over other LLM-based verifiers (discussed in \Cref{subsec:comparison}) establishes it as an appropriate framework for evaluating LLMs' invariant generation capabilities.

We provide further analysis of LLM inference time in Appendix~\ref{subsec:inference_time}, the discussion of the trade-off between solving time and LLM inference cost in Appendix~\ref{subsec:tradeoff}, and evaluate alternative prompting strategies (few-shot and chain-of-thought prompting) in Appendix~\ref{subsec:prompting}.

\subsection{Failure Mode Analysis}
\label{subsec:failure-analysis}

To understand why many invariants fail to accelerate verification in \name{}, we analyze the failure modes of the top three models, with detailed results provided in Appendix~\ref{subsec:falueappendix}. The analysis shows that although some failures are caused by incorrect invariants, the majority are due to timeouts.

We randomly sampled 50 timeout cases and manually analyzed the invariants. Our analysis reveals several key patterns. First, only 7 cases (14\%) involved genuinely incorrect invariants, indicating reasonable LLM correctness. Second, 19 cases (38\%) generated invariants identical to the original assertion, providing no decomposition benefit, with 4 additional cases nearly identical. Third, 9 cases (18\%) generated trivial invariants (e.g., simple bounds) that failed to capture essential loop behavior. Fourth, only 5 cases (10\%) produced strong, non-trivial invariants that still timed out, suggesting inherently difficult verification queries. Finally, 6 cases are incorrect original assertions where the LLM generated correct invariants. We provide case studies in Appendix~\ref{subsec:timeout-case-study}.


\subsection{Results of Fine-Tuning and Best-of-N Sampling}
\label{subsec:sft}

\begin{wrapfigure}{r}{0.55\linewidth}
    \centering
    \includegraphics[width=\linewidth]{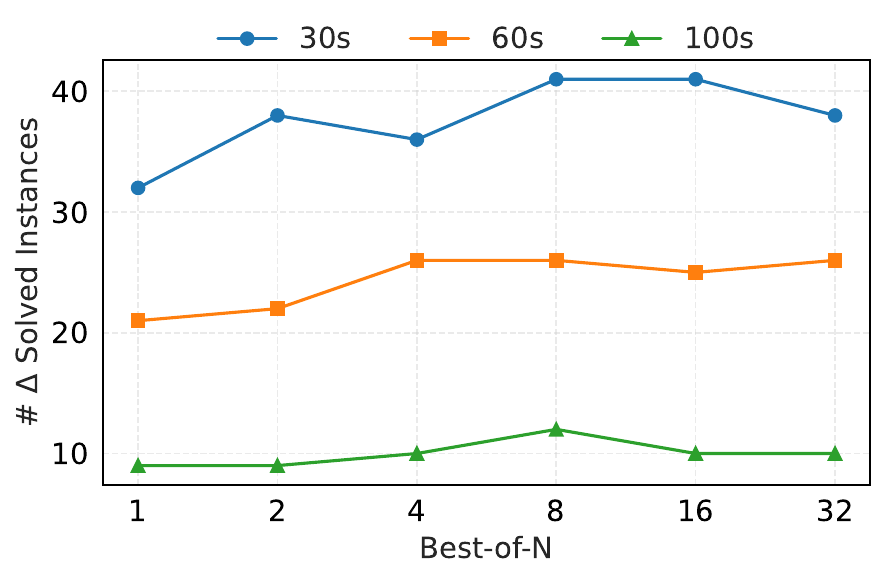}
    \vspace{-2em}
    \caption{Effect of Best-of-N sampling on the number of $\Delta$ instances solved over Uautomizer under different timeout settings.}
    \vspace{-1em}
    \label{fig:bestofn}
\end{wrapfigure}

\paragraph{Modest Improvement from Supervised Fine-Tuning.}
We perform supervised fine-tuning using LoRA~\citep{hu2022lora} on Qwen2.5-7B for 3 epochs. As shown in \Cref{tab:sft}, fine-tuning yields modest improvements: verified invariants increase from 419 to 431, and additionally solved instances increase from 5 to 7 under 60s timeout. These results suggest that domain-specific fine-tuning can improve both correctness and speedup, though gains remain moderate. The broader value of this direction is also reflected in subsequent work that builds on UAutomizer-generated invariants for fine-tuning and investigates whether stronger data curation can improve verification acceleration~\citep{pinto2026not}.

\paragraph{Benefits of Best-of-N Sampling.}
We evaluate Best-of-N sampling by generating multiple invariant candidates from gpt-5.2 at temperature 0.7 and selecting the one yielding the fastest verification time. As shown in \Cref{fig:bestofn}, increasing N from 1 to 4 generally improves performance. The benefits continue to grow up to N=8, which achieves the best results. Beyond N=8, performance plateaus as resource contention and memory constraints begin to offset the advantages of additional sampling. These results demonstrate that Best-of-N sampling is an effective strategy for improving verification performance, with optimal gains observed at N=8 in our experimental setup.

\subsection{Comparison with Prior LLM-based Verifiers}
\label{subsec:comparison}

\begin{table*}[!tb]
  \centering
  \small
  \renewcommand{\arraystretch}{1.2}
  \setlength{\tabcolsep}{5pt}
  \begin{tabular}{p{3.8cm}cccccccccc}
    \toprule
    \multirow{2}{*}{\textbf{Method}}
      & \multicolumn{2}{c}{\textbf{30s}}      & \multicolumn{2}{c}{\textbf{60s}}      & \multicolumn{2}{c}{\textbf{250s}}      & \multicolumn{2}{c}{\textbf{350s}}      & \multicolumn{2}{c}{\textbf{500s}} \\
    \cmidrule(lr){2-3} \cmidrule(lr){4-5} \cmidrule(lr){6-7} \cmidrule(lr){8-9} \cmidrule(lr){10-11}
      & \# $\Delta$ & $\overline{T}$(s) & \# $\Delta$  & $\overline{T}$(s) & \# $\Delta$  & $\overline{T}$(s) & \# $\Delta$ & $\overline{T}$(s) & \# $\Delta$  & $\overline{T}$(s) \\
    \midrule
    LaM4Inv & 1 & 30.0 & 1 & 59.9 & 1 & 249.3 & 0 & 348.9 & 0 & 498.4 \\
    Clause2Inv-original & 14 & 25.8 & 9 & 49.9 & 4 & 199.2 & 2 & 277.4 & 0 & 394.4 \\
    Clause2Inv-gpt-5.2 & 20 & 25.1 & 13 & 48.2 & 6 & 190.7 & 3 & 265.2 & 1 & 377.0 \\
  Loopy & 0 & 30.0 & 0 & 60.0 & 5 & 240.0 & 3 & 299.5 & 2 & 360.4 \\
    LEMUR-original & 7 & 28.6 & 7 & 50.4 & 3 & 165.3 & 1 & 222.3 & 1 & 306.8 \\
    LEMUR-gpt-5.2 & 15 & 27.9 & 13 & 47.9 & 7 & 153.9 & 4 & 205.9 & 2 & 282.8 \\
    \textbf{Quokka-gpt-5.2} (best-of-8) & 41 & 17.5 & 25 & 26.6 & 7 & 78.5 & 4 & 104.8 & 3 & 143.8 \\
    \textbf{Quokka-gpt-5.2}  (2-shot) & 33 & 14.3 & 20 & 19.4 & 6 & 45.2 & 2 & 57.6 & 2 & 76.0 \\
    \bottomrule
  \end{tabular}
  \label{tab:extra_vs_uautomizer}
  \caption{Extra solved instances over UAutomizer and average solving time at different timeout thresholds.}
\end{table*}

\begin{wrapfigure}{r}{0.55\linewidth}
    \centering
    \vspace{-0.5\baselineskip}
    \includegraphics[width=\linewidth]{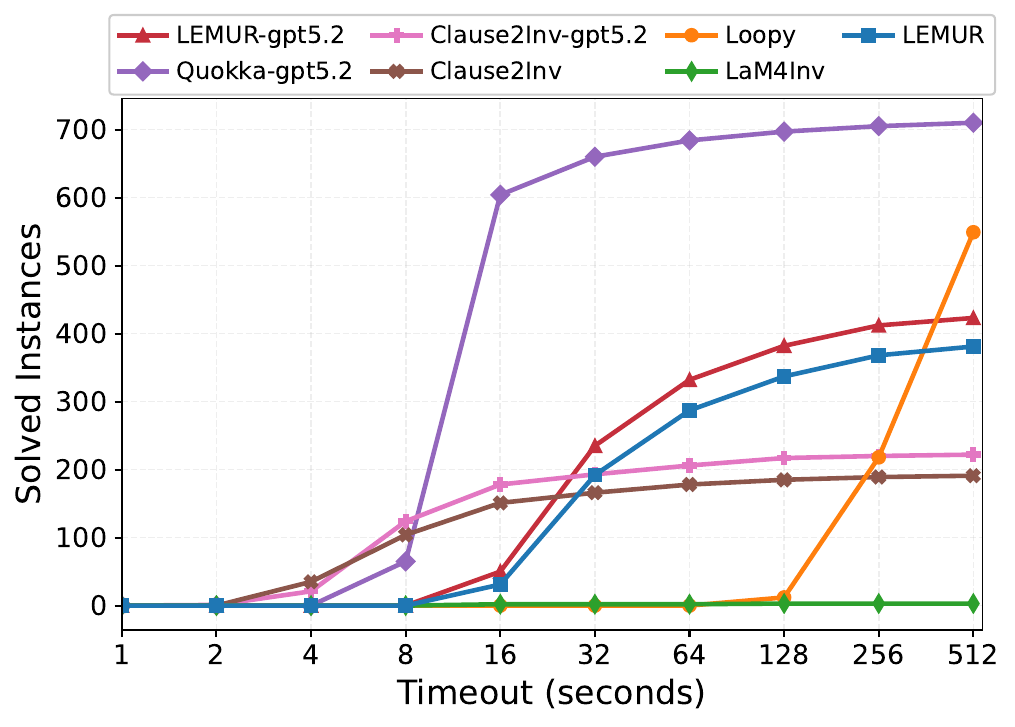}
    \vspace{-2em}
    \caption{Number of instances solved by different methods over varying timeout budgets.}
     \vspace{-1em}
    \label{fig:comparison}
\end{wrapfigure}

\Cref{tab:extra_vs_uautomizer} compares \name{} against prior LLM-based verifiers, reporting additional instances solved beyond \uauto~and average solving time under different timeout budgets. \Cref{fig:comparison} shows cumulative solved instances over time, demonstrating \name{}'s superior performance across all timeout thresholds.

\begin{table*}[!tb]
  \centering
\small
  \renewcommand{\arraystretch}{1.2}
  \setlength{\tabcolsep}{2.9pt}
  \begin{tabular}{p{3cm}cccccccccc}
    \toprule
    \multirow{2}{*}{\textbf{Method}}
      & \multicolumn{2}{c}{\textbf{30s}}
      & \multicolumn{2}{c}{\textbf{60s}}
      & \multicolumn{2}{c}{\textbf{250s}}
      & \multicolumn{2}{c}{\textbf{350s}}
      & \multicolumn{2}{c}{\textbf{500s}} \\
    \cmidrule(lr){2-3} \cmidrule(lr){4-5} \cmidrule(lr){6-7} \cmidrule(lr){8-9} \cmidrule(lr){10-11}
      & \# Solved & $\overline{T}$(s)
      & \# Solved & $\overline{T}$(s)
      & \# Solved & $\overline{T}$(s)
      & \# Solved &$\overline{T}$(s)
      & \# Solved & $\overline{T}$(s) \\
    \midrule
    LEMUR-gpt-5.2 & 4 & 27.9 & 9 & 34.0 & 18 & 75.9 & 19 & 88.1 & 20 & 106.6 \\
    \textbf{Quokka-gpt-5.2} & 19 & 9.7 & 20 & 11.6 & 24 & 28.8 & 24 & 28.8 & 24 & 28.8 \\
        \bottomrule
  \end{tabular}
  \label{tab:lemur_comparison}
    \caption{Performance comparison on the benchmark used in LEMUR. Since \uauto~solves 0 instances within ten minutes on this dataset, \# Solved equals \# $\Delta$ (additional instances solved beyond the baseline). Results shown for different timeout thresholds.}
\end{table*}

\paragraph{\name{} Outperforms Prior LLM-based Verifiers.} \name{} with gpt-5.2 and best-of-8 sampling achieves state-of-the-art results across all speedup metrics, while \name{} with gpt-5.2 and 2-shot prompting also achieves leading performance under most timeout thresholds. Among prior work, LEMUR~\citep{wu2024lemur} and Clause2Inv~\citep{cao2025clause2inv} represent the strongest baselines. LEMUR supports both UAutomizer and ESBMC. We evaluate LEMUR with UAutomizer only, which is the state-of-the-art solver and matches the base solver used in our method. For fair comparison, we upgrade both to gpt-5.2 (from gpt-4 and gpt-4o-mini respectively), creating LEMUR-gpt-5.2 and Clause2Inv-gpt-5.2. \name{}-gpt-5.2 outperforms both upgraded baselines. LaM4Inv~\citep{wu2024llm} and LOOPY~\citep{kamath2024leveraging} show minimal gains over \uauto. LaM4Inv's performance differs from its original paper due to dataset differences and its requirement for manual preprocessing (confirmed by the authors), for which no automated script exists.

To further quantify the performance gap at larger timeout, we evaluate on LEMUR's benchmark (\Cref{tab:lemur_comparison})—challenging instances where \uauto~fails within 10 minutes. \name{} consistently outperforms LEMUR-gpt-5.2 across all timeouts, solving substantially more instances with lower average solving times.

\paragraph{Analysis.} The baselines differ in both base solver choices and algorithm design. We analyze base verifier differences in Appendix~\ref{subsec:app:solvercompare}. Regarding algorithm design, \name{} provides a clean interface between an LLM and a verifier, representing a conceptual shift from prior work that introduces complex algorithms to transform imperfect LLM outputs into usable invariants. Prior approaches treat LLM output as noisy symbolic material requiring algorithmic repair, whereas \name{} treats it as a direct proof candidate to be validated by the verifier. For instance, LaM4Inv requests complete invariants but then filters predicates with a model checker and reassembles them, indicating distrust in the raw output. Clause2Inv takes this further. However, as models become increasingly capable, simpler evaluation-centric designs become preferable. We additionally provide a detailed comparison between \name{} and LEMUR in Appendix~\ref{subsec:app:solvercompare}.
\section{Discussion}

\paragraph{Run-to-Run Variance.}
We repeat the experiments in Tables~2--4 and Figure~2 with three random seeds and observe consistent trends. In the SFT experiment, the number of instances solved within 60 seconds increases from \(400 \pm 5.9\) to \(413 \pm 6.1\) after fine-tuning, with 95\% confidence intervals. Best-of-N continues to outperform single-sample generation and saturates around \(N=8\), while GPT-5.2 remains the strongest zero-shot model and Quokka-gpt-5.2 with Best-of-8 remains competitive with prior LLM-based verifiers.

\paragraph{Effect of Candidate Invariant Locations.}
We ablate the loop-beginning restriction by allowing the LLM to choose among the location types used by LEMUR. Under the 500-second budget, the original setting solves 710 instances versus 697 with expanded locations, suggesting that the simpler restriction does not hurt performance.

\paragraph{Effect of Longer Verification Timeouts.}
We further evaluate GPT-5.2 under larger verification time budgets. Quokka solves 710 instances under the original time budget of 500 seconds, 761 under 1200 seconds, and 763 under 1800 seconds. These results show that additional verification time yields a noticeable gain by 1200 seconds, while the gains largely saturate by 1800 seconds.
\section{Conclusion}
\label{sec:conclusion}

This work introduced \name{}, a framework for LLM-based invariant synthesis that accelerates program verification with soundness guarantees and state-of-the-art performance. Unlike prior work that treats LLM outputs as noisy symbolic material requiring substantial post-processing, \name{} adopts a simpler design that directly validates whether each LLM-generated invariant helps prove the target assertion. We constructed a benchmark of \evalsize{} instances derived from SV-COMP and evaluated \numllm{} LLMs across multiple model families. Our results demonstrate that \name{} consistently outperforms prior LLM-based verifiers. We further showed that supervised fine-tuning and Best-of-N sampling yield measurable improvements in accelerating verification. Our work establishes a new state-of-the-art for LLM-based program verification and provides a principled foundation for future research in this direction.

\section*{Acknowledgments}
This work was supported in part by a Google Research Award. We thank Clark Barrett for the discussions.

\bibliography{colm2026_conference}
\bibliographystyle{colm2026_conference}

\setcounter{figure}{0}
\renewcommand{\thefigure}{A\arabic{figure}}
\setcounter{table}{0}
\renewcommand{\thetable}{A\arabic{table}}

\newpage
\appendix
\onecolumn
\section{Appendix}
\label{sec:app}

\subsection{Additional Details on Experiment Setup}
\label{subsec:app:setup}

\paragraph{Models.}
We benchmark Claude models from Anthropic, GPT models from OpenAI, and models from the Qwen and LLaMA families~\citep{hui2024qwen2,yang2025qwen3,touvron2023llama}.

\paragraph{Hardware and OS.} Experiments were conducted on a node with Intel E5-2640 v4 CPUs, 256G main memory, running Ubuntu 20.04 LTS.

\subsection{Prompt Template Used for Training Dataset Generation}
\label{subsec:prompt-template}

\Cref{tab:prompt} shows the prompt template used for synthesizing training programs from seed programs.

\begin{table}[ht]
\begin{tcolorbox}[width=\linewidth]
{
You will be shown 3 example C programs. Please gain inspiration from the following programs to create a new high-quality C program. Do not simply copy from any of them. \\

Requirements for the generated program:\\
1. The program MUST contain non-trivial loops (for or while).\\
2. The program MUST contain assertions.\\
3. The program MUST be compilable, self-contained, and reasonably complex (not trivial or overly short).\\
4. Only output the new C program.\\

Example snippets:\\
\textbf{Program 1}:\\
\texttt{\{SEED\_PROGRAM\_1\}}\\[2pt]
\textbf{Program 2}:\\
\texttt{\{SEED\_PROGRAM\_2\}}\\[2pt]
\textbf{Program 3}:\\
\texttt{\{SEED\_PROGRAM\_3\}}\\

Output format: The generated program must be wrapped strictly in the following format:\\
\verb|```c|\\
\texttt{<NEW\_C\_PROGRAM>}\\
\verb|```|
}
\end{tcolorbox}
\caption{Prompt template for synthetic data generation.}
\label{tab:prompt}
\end{table}

\subsection{Prompt Template Used for Invariant Generation}
\label{subsec:invariant-prompt-template}

\Cref{tab:invariant-prompt} shows the prompt template used to prompt the LLM to generate a loop invariant for a given C program.

\begin{table}[ht]
\begin{tcolorbox}[width=\linewidth]
{
\textbf{System prompt:}\\
\texttt{Given a C program with line numbers, you need to identify the loop invariant to be added at one of the provided points in the program. A loop invariant is an expression that is true before, during, and after the execution of a loop.}\\

\texttt{You will be provided the set of points in the program where loop invariants can be added. Each point will be specified as after line X.}\\

\texttt{Choose one of the points to insert a loop invariant. Respond with exactly this format:}\\
\texttt{"After line X, insert assume(condition);"}\\

\texttt{Where:}\\
\texttt{- X refers to the provided line number where the loop invariant needs to be added}\\
\texttt{- condition is the loop invariant condition you want to insert. The loop invariant should not modify any variable (no ++, --, +=, -=, =).}\\
\texttt{- only output one loop invariant}\\[4pt]

\textbf{User prompt:}\\
\texttt{Program:}\\
\texttt{\{PROGRAM\}}\\[2pt]

\texttt{Choose one of the points to insert a correct loop invariant at in the program:}\\
\texttt{\{POINTS\}}\\[2pt]

\texttt{Please analyze the provided C program and output only where to insert a loop invariant in the specified format.}
}
\end{tcolorbox}
\caption{Prompt template used to generate a loop invariant for a given program. The placeholder \texttt{\{PROGRAM\}} denotes the input C program with line numbers, and \texttt{\{POINTS\}} denotes the candidate locations where the invariant may be inserted.}
\label{tab:invariant-prompt}
\end{table}

\subsection{An Example from the Fine-Tuning Dataset}
\label{subsec:sftexample}

\begin{figure}[!tb]
  \centering
  \noindent\textbf{Original Program (Input to \uauto)}
\begin{lstlisting}
int main() {
  int N = __VERIFIER_nondet_int();
  assume_abort_if_not(N >= 1 && N < 100);
  int x = 0, y = 0;
  for (int i = 0; i < N; i++) {
      if (i % 2 == 0) {
          x += i;
      } else {
          y += i;
      }
  }
  int diff = x - y;
  __VERIFIER_assert(diff <= N);
  return 0;
}
\end{lstlisting}

  \vspace{0.5em}
  \noindent\textbf{Loop Invariant Generated by \uauto:}
\begin{lstlisting}
Line Number: 6
Predicate:
(
(i + 1) % 2 == 0 &&
x < 2 + i + y &&
x < 2 + y + 2 * i &&
x < N + y + 1 &&
1 <= N
)
||
(
i % 2 == 0 &&
x < 2 + i + y &&
x < 2 + y &&
1 <= N
)
\end{lstlisting}
  \caption{An example from the fine-tuning dataset: program and its loop invariant generated by \uauto.}
  \label{fig:uautomizer-example}
\end{figure}

\Cref{fig:uautomizer-example} shows an example from the fine-tuning dataset with the program to the \uauto and the generated loop invariant.

The loop invariant holds at Line~6 (the beginning of the loop).
It is a disjunction of two clauses, and can be written as $I \equiv P \lor Q = \big( (i+1)\bmod 2 = 0 \land x < 2 + i + y \land x < 2 + y + 2i \land x < N + y + 1 \land 1 \le N \big)
\lor \big( i \bmod 2 = 0 \land x < 2 + i + y \land x < 2 + y \land 1 \le N \big)$.

By inspecting the loop, we can derive an exact relationship between
$x$, $y$, and $i$ at the beginning of each iteration.
Since $x$ accumulates all even numbers less than $i$ and $y$ accumulates all odd numbers less 
than $i$, we obtain:
\[
\text{if $i$ is even:}\quad x - y = -\frac{i}{2},
\qquad
\text{if $i$ is odd:}\quad x - y = \frac{i - 1}{2}.
\]

Using this relationship, it is straightforward to verify that the invariant $I$ produced by 
\uauto is correct. Moreover, $I$ is strong enough to prove the final assertion 
$x - y \le N$.

To show that the assertion holds at loop termination, we can check the following verification condition
\[
I \;\wedge\; (i = N) \;\Rightarrow\; (x - y \le N),
\]
where $i = N$ denotes the loop’s exit condition.

\paragraph{Case 1: $N$ is odd.}
Instantiating the invariant with $i = N$ activates the $P$ disjunct of~$I$, from which we 
obtain
\[
x - y < N + 1.
\]
Since $x - y$ is an integer, this directly implies $x - y \le N$.

\paragraph{Case 2: $N$ is even.}
In this case, the $Q$ disjunct applies. From the clause $x < 2 + y$ contained in $Q$, we 
derive
\[
x - y < 2.
\]
Because $x - y$ is an integer and $N \ge 1$ and even (hence $N \ge 2$), we conclude
\[
x - y \le 1 \le N,
\]
establishing the desired post-condition.

Thus, the invariant $I$ indeed suffices to prove the final assertion.

\subsection{Results of \uauto on Prior Datasets}
\label{subsec:resultprior}

\begin{table}[!tb]
    \centering
    \small
\begin{tabular}{l|c|*{5}{p{1.4cm}<{\centering}}}
\toprule
 \multirow{2}{*}{\textbf{Verifier}} & \multirow{2}{*}{\textbf{Total Instances}} & \multicolumn{4}{c}{\textbf{Solved Instances under Different Timeouts}} \\
\cmidrule(lr){3-6}
 &  & \textbf{10s} & \textbf{100s} & \textbf{300s} & \textbf{600s} \\
\midrule
LaM4Inv~\citep{wu2024llm} & 316 & 144 & 286 & 295 & 299 \\
\uauto  & 316 & 299 & 299 & 299 & 299 \\
\midrule
Loopy~\citep{kamath2023finding} & 469 & 0 & 133 & 353 & 403 \\
\uauto  & 469 & 372 & 403 & 411 & 413 \\
\midrule
LEMUR~\citep{wu2024lemur} & 47 & 2 & 8 & 16 & 19 \\
\uauto  & 47 & 0 & 0 & 0 & 0 \\
\bottomrule
\end{tabular}
\vspace{0.5em}
\caption{Comparison of prior LLM-based verifiers and \uauto on their own custom dataset under different timeout budgets.}
\label{tab:uauto_prior}
\end{table}

As shown in Table~\ref{tab:uauto_prior}, \uauto, the state-of-the-art non-LLM-based verifier, consistently solves more instances than LaM4Inv~\citep{wu2024llm} and Loopy~\citep{kamath2023finding}, two representative LLM-based tools, on their respective custom datasets. This demonstrates the importance of base solver choice.

LEMUR~\citep{wu2024lemur} is the only tool that surpasses \uauto on its own benchmark. This outcome is largely explained by LEMUR's dataset construction methodology: it specifically curates problems that \uauto cannot solve within 600 seconds. To assess whether LEMUR's advantage generalizes beyond this carefully selected benchmark, we conduct an additional analysis. We randomly selected 50 instances that \uauto failed to solve within 600 seconds and evaluated LEMUR's performance on them. LEMUR successfully solved 12 of these 50 instances within the same 600-second timeout, demonstrating that its gains are not merely an artifact of benchmark design but reflect genuine capability on challenging problems. Nevertheless, for our work, we focus on problems that \uauto can solve within the time budget, as our primary objective is to quantify speedup.

\subsection{Inference Time of Different Models}
\label{subsec:inference_time}

\Cref{tab:inference_time} presents the average inference time per instance for different LLMs when generating invariants. As expected, smaller open-source models such as Llama-3.1-8B and Qwen2.5-7B exhibit the fastest generation times, averaging around 0.3--0.4 seconds per instance. Mid-sized models like Qwen2.5-72B and the GPT-5 series require approximately 0.6--1.4 seconds. The Claude models take longer at around 1.6--2.7 seconds per instance. The o3 model has the highest inference overhead, averaging over 8 seconds per instance. These results highlight the trade-off between model capability and inference cost, which must be considered when evaluating end-to-end verification performance.

\begin{table*}[!tb]
    \centering
    \small
    \begin{tabular}{lc}
    \toprule
    \textbf{Model} & \textbf{Avg Generation Time (s)} \\
    \midrule
    Llama-3.1-8B      &      0.300  \\
    Qwen2.5-7B             &      0.355  \\
    Qwen2.5-72B            &      0.578  \\
    gpt-5.2                                  &      1.076  \\
    gpt-5.1                       &      1.373  \\
    claude-opus-4.1            &      1.626  \\
    claude-opus-4.5               &      2.538  \\
    claude-sonnet-4           &      2.746  \\
    o3                   &      8.075  \\
    \bottomrule
    \end{tabular}
    \caption{Average inference time per instance for different LLMs when generating invariants.}
    \label{tab:inference_time}
\end{table*}

\begin{table*}[!tb]
  \centering
  \small
    \renewcommand{\arraystretch}{1.2}
  \setlength{\tabcolsep}{5.5pt}
  \renewcommand{\arraystretch}{1.2}
  \setlength{\tabcolsep}{5.5pt}
  \begin{tabular}{lcccccccc}
\toprule
\multirow{2}{*}{\textbf{Model}} & \multirow{2}{*}{\textbf{\# Verified Invariants}} & \multicolumn{3}{c}{\textbf{60s}} & \multicolumn{3}{c}{\textbf{500s}} \\
\cmidrule(lr){3-5} \cmidrule(lr){6-8}
& & $\Delta$ & \# Solved & $\overline{T}$ & $\Delta$ & \# Solved & $\overline{T}$ \\
\midrule
 gpt-5.1& 399 & 31 & 394 & 34.8 & 1 & 399 & 272.9 \\
gpt-5.2 & 455 & 30 & 450 & 31.2 & 1 & 455 & 240.9 \\

    \bottomrule
  \end{tabular}
  \caption{Performance of different LLMs when using ESBMC as the base verifier.}
  \label{tab:esbmc}
\end{table*}

\subsection{Results for ESBMC as the Base Solver}
\label{subsec:esbmc}

To demonstrate the generality of \name{}'s methodology, we conduct additional experiments using ESBMC~\citep{gadelha2018esbmc} as the base solver instead of \uauto. As shown in \Cref{tab:esbmc}, \name{} successfully accelerates ESBMC as well, with both gpt-5.1 and gpt-5.2 generating hundreds of verified invariants and solving additional instances. These results confirm that \name{} is not limited to a specific solver and can be applied to accelerate different verification tools.

\subsection{Analysis of Solving Time and LLM Inference Cost}
\label{subsec:tradeoff}

\begin{figure}[!tb]
    \centering
    \includegraphics[width=0.9\linewidth]{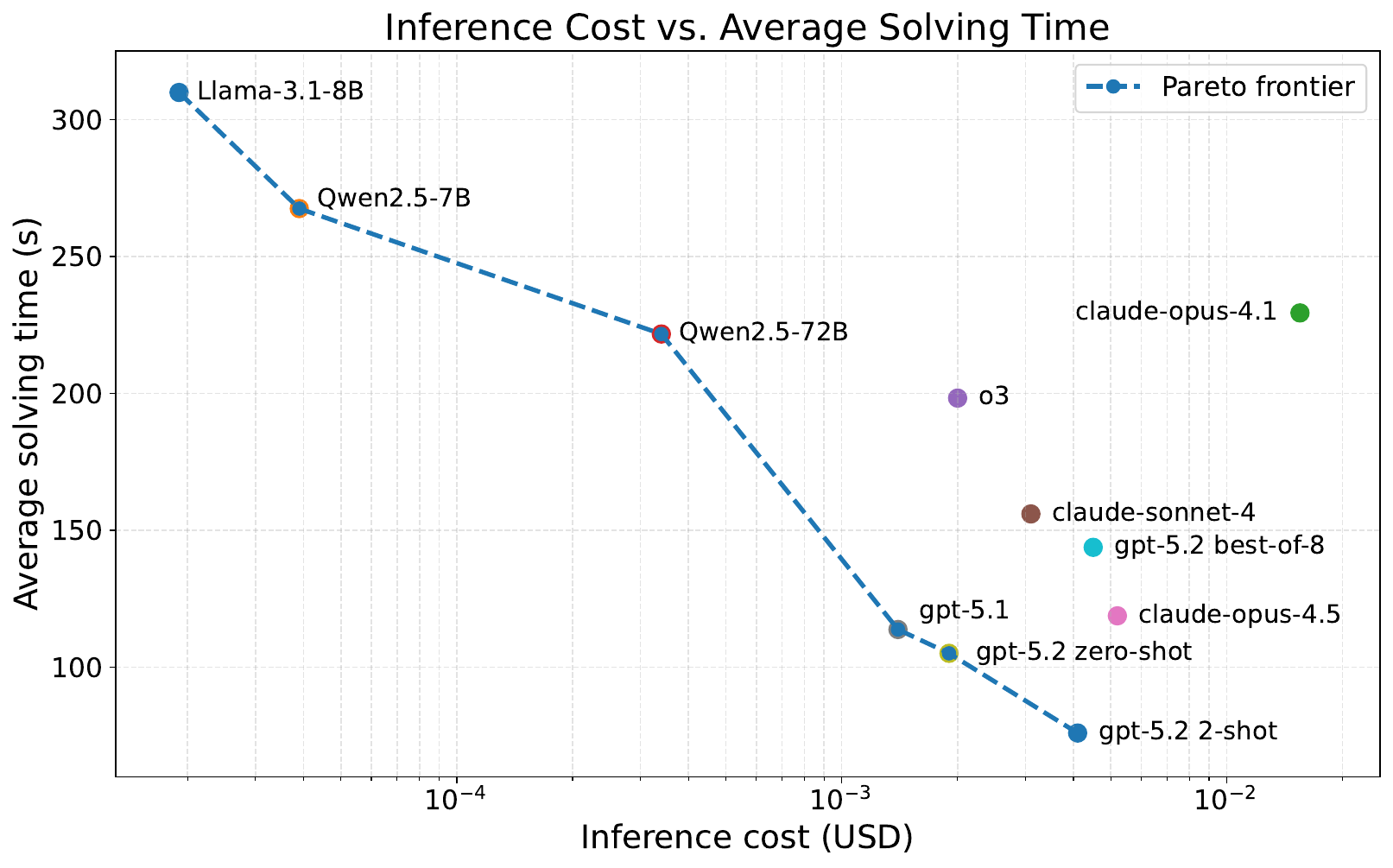}
    \caption{
    Trade-off between average solving time and LLM inference cost per instance across model and prompting configurations. Lower values on both axes are preferable.}
    \label{fig:pareto}
\end{figure}

\Cref{fig:pareto} plots the average solving time against the LLM inference cost per instance, where configurations toward the lower left are preferable. We identify the Pareto frontier as the set of configurations for which no alternative achieves both lower inference cost and lower solving time.

\subsection{Results for Prompting-Based Methods}
\label{subsec:prompting}

\begin{table*}[!tb]
  \centering
\small
  \renewcommand{\arraystretch}{1.2}
  \setlength{\tabcolsep}{5pt}
    \renewcommand{\arraystretch}{1.2}
  \setlength{\tabcolsep}{5pt}
  \begin{tabular}{lcccccccc}
  \toprule
    \multirow{2}{*}{\textbf{Model}} & \multirow{2}{*}{\textbf{\# Verified Invariants}} & \multicolumn{3}{c}{\textbf{60s}} & \multicolumn{3}{c}{\textbf{500s}} \\
    \cmidrule(lr){3-5} \cmidrule(lr){6-8}
    & & \# $\Delta$ & \# Solved & $\overline{T}$ & \# $\Delta$ & \# Solved & $\overline{T}$ \\
    \midrule
    gpt-5.2 (CoT) & 639 & 18 & 609 & 33.4 & 2 & 638 & 153.5 \\
    gpt-5.2 & 710 & 21 & 681 & 22.2 & 1 & 710 & 105.1 \\
    gpt-5.2 (4-shot) & 756 & 20 & 721 & 20.1 & 0 & 754 & 81.1 \\
    gpt-5.2 (2-shot) & 764 & 20 & 730 & 19.4 & 2 & 763 & 76.0 \\

    \bottomrule
  \end{tabular}
  \caption{Comparison of different prompting strategies for gpt-5.2.}
  \label{tab:prompting}
\end{table*}

We investigate the impact of different prompting strategies on invariant generation quality using gpt-5.2. As shown in \Cref{tab:prompting}, we compare chain-of-thought (CoT) prompting, few-shot prompting with 2 and 4 examples, and our default zero-shot approach. The results show that few-shot prompting can sometimes improve performance (e.g., 2-shot is generally better than 0-shot), but more examples may also introduce noise or distract the model from the task at hand (e.g., 4-shot is worse than 2-shot). Chain-of-thought prompting yields the weakest results across all metrics, with only 639 verified invariants and the fewest solved instances. This suggests that explicit reasoning steps may increase generation time without proportional benefits to invariant quality, thereby reducing overall speedup. These findings indicate that few-shot prompting could improve the task of invariant synthesis.

\subsection{Detailed Comparison against Prior Work}
\label{subsec:app:solvercompare}

\paragraph{Comparison of Base Solver} Different LLM-based verification frameworks are built on different base solvers: \name{} and LEMUR use \uauto, LOOPY uses Frama-C~\citep{cuoq2012frama}, and LaM4Inv uses ESBMC~\citep{gadelha2018esbmc}. The choice of base solver significantly impacts the performance, as stronger solvers provide a better foundation for speedup. Since \uauto represents the current state-of-the-art solver, we believe future work should prioritize developing LLM-based verifiers atop state-of-the-art solvers such as \uauto. For a direct comparison between \uauto and prior LLM-based solvers, we also refer readers to Appendix~\ref{subsec:resultprior}, where we evaluate \uauto on the dataset released by prior works.

\paragraph{Detailed Comparison with LEMUR.} To better understand the performance differences between LEMUR and \name{}, we analyze their implementation strategies and examine the instances where \name{} outperforms LEMUR. LEMUR allows LLMs to insert invariants at multiple locations (both before and at the beginning of loops) and generates multiple candidate invariants for each location. It then sequentially evaluates each candidate, verifying both the original assertion and the invariant itself. If verification times out within 30 seconds, LEMUR \emph{recursively} invokes the LLM again to generate additional invariants. In contrast, \name{} adopts a much simpler strategy: it restricts invariant placement to the beginning of loops, asks the LLM to select a single location (when multiple loops exist), and generates only one invariant. The two verification queries (for the assertion and the invariant) are then executed in parallel without recursion or additional LLM invocations. Among the 353 instances that \name{} solves but LEMUR does not, 129 instances (36.5\%) contain multiple loops. We hypothesize that for programs with multiple loops, LEMUR's exhaustive search strategy becomes less effective due to the combinatorially larger space of invariant placements and candidates it must explore. In contrast, \name{}'s focused approach (i.e., requiring the LLM to commit to a single location and invariant) appears more effective in these complex scenarios. We identify three key advantages of our design. First, parallel execution of verification queries reduces wall-clock time by simultaneously checking both the assertion and the invariant. Second, constraining the LLM to select a single location and invariant forces more deliberate reasoning about which invariant will be most impactful, rather than relying on exhaustive search over many candidates. Third, eliminating recursion and repair loops reduces both implementation complexity and the risk of execution failures (which we observe in LEMUR's logs). Together, these design choices enable \name{} to achieve superior performance while maintaining a simpler and more robust implementation.



\begin{table}[!tb]
    \centering
    \small
    \begin{tabular}{lcccc}
        \toprule
        \textbf{Model} & \textbf{\# Incorrect} & \textbf{\# Goal Timeout} & \textbf{\# Invariant Timeout} & \textbf{\# Both Timeout} \\
        \midrule
        gpt-5.1  & 78 & 96 & 245 & 110 \\
        claude-sonnet-4 & 110 & 160 & 209 & 145 \\
        gpt-5.2           & 56 & 139  & 230 & 121 \\
        \bottomrule
    \end{tabular}
    \vspace{0.5em}
    \caption{Failure mode breakdown for the top three models on \name{}. \emph{Incorrect}: the invariant is refuted by the solver. \emph{Goal Timeout}: the invariant is verified, but proving the assertion under it times out. \emph{Invariant Timeout}: verifying the invariant itself times out. \emph{Both Timeout}: both verification queries time out.}
    \label{tab:failure}
\end{table}

\subsection{Failure Mode Analysis}
\label{subsec:falueappendix}

As shown in \Cref{tab:failure}, we categorize failures into four types: 1) \emph{Incorrect Invariants}: the invariant is refuted by the solver; 2) \emph{Goal Timeout}: the invariant is verified, but proving the assertion under it times out; 3) \emph{Invariant Timeout}: verifying the invariant itself times out; 4) \emph{Both Timeout}: both verification queries time out.

\subsection{Case Studies: Verification Timeout in Quokka}
\label{subsec:timeout-case-study}

To better understand the verification timeouts in Quokka, we manually examined 50 timeout cases and identified five recurring categories. In this section, we select two of these categories for closer analysis and present one representative program from each. The first category includes cases where the LLM proposes a trivial but correct invariant, yet the solver still cannot complete verification within the timeout budget. The second category includes cases where the LLM proposes a correct invariant that is almost identical to the original assertion.

\paragraph{Case 1: Trivial and correct invariant.}
\Cref{fig:timeout-case-weak} shows a benchmark where the LLM proposes the invariant \texttt{c == y} after line 12. This invariant is correct because \texttt{c} and \texttt{y} are both initialized to zero before the loop and are incremented by exactly one in every iteration, so the equality is preserved throughout the loop. Moreover, it is trivial to prove: its validity follows directly from the initialization and the loop updates, and can be established by a simple inductive argument. Nevertheless, in this case, the solver still cannot prove this invariant within the timeout budget. This example shows that verification timeout may occur even when the proposed invariant is both semantically correct and trivial to prove.

\begin{figure}[!tb]
  \centering
\begin{lstlisting}
int main() {
    short k;
    long long y, x, c;
    k = __VERIFIER_nondet_short();
    assume_abort_if_not(k >= 0 && k <= 5);
    assume_abort_if_not(k <= 256);

    y = 0;
    x = 0;
    c = 0;

    while (1) {
        __VERIFIER_assert(-2 * y * y * y * y * y * y - 6 * y * y * y * y * y - 5 * y * y * y * y + y * y + 12 * x == 0);
        
        if (!(c < k)) {
            break;
        }
        c = c + 1;
        y = y + 1;
        x = y * y * y * y * y + x;
    }

    return 0;
}
\end{lstlisting}
  \caption{A timeout case where the proposed invariant \texttt{c == y} is semantically correct and trivial to prove, yet the solver still cannot establish it within the timeout budget.}
  \label{fig:timeout-case-weak}
\end{figure}

\paragraph{Case 2: Correct invariant as a trivial variant of the original assertion.}
A different phenomenon is shown in \Cref{fig:timeout-case-nearpost}, where the LLM proposes the invariant
\[
x = \frac{y(y+1)}{2}.
\]
This invariant is correct: it holds initially and is preserved by the loop updates. However, it is essentially just a trivial algebraic reformulation of the original assertion
\[
y^2 - 2x + y = 0.
\]
This possibly suggests a limitation of the LLM in decomposing the verification task into a genuinely simpler intermediate property, instead of merely restating the target assertion in an equivalent form.

\begin{figure}[!tb]
  \centering
\begin{lstlisting}
int main() {
    int k;
    long long y, x, c;
    k = __VERIFIER_nondet_int();
    assume_abort_if_not(k >= 0 && k <= 10);

    y = 0;
    x = 0;
    c = 0;

    while (1) {

        if (!(c < k)) {
            break;
        }

        c = c + 1;
        y = y + 1;
        x = y + x;
    }
    __VERIFIER_assert((y * y) - 2 * x + y == 0);
    return 0;
}


\end{lstlisting}
  \caption{A case where the proposed invariant is correct but is essentially a trivial algebraic variant of the original assertion.}
  \label{fig:timeout-case-nearpost}
\end{figure}

\end{document}